\documentclass[a4paper,11pt]{article}
\pdfoutput=1 
\usepackage{aas_macros}
\usepackage{jcappub} 
\usepackage[T1]{fontenc} 
\usepackage{amsbsy}
\usepackage{color}
\usepackage{xcolor}
\usepackage{subcaption}
\usepackage{hyperref}

\usepackage[normalem]{ulem}

\newcommand{\be}{\begin{equation}}
\newcommand{\ee}{\end{equation}}

\usepackage{amssymb}

\title{\boldmath Hydrodynamical simulations of galaxy formation with non-Gaussian initial conditions}

\author[a]{Cl\'ement Stahl,}
\author[b]{Yohan Dubois,}
\author[a]{Benoit Famaey,}
\author[c]{Oliver Hahn}
\author[a]{Rodrigo Ibata,}
\author[a]{Katarina Kraljic,}
\author[c]{Thomas Montandon,}

\affiliation[a]{Universit\'e de Strasbourg, CNRS, Observatoire astronomique de Strasbourg, UMR 7550, 67000 Strasbourg, France}
\affiliation[b]{Institut d'Astrophysique de Paris, CNRS, Sorbonne Universit\'{e}, UMR 7095, 98bis bd Arago, 75014 Paris, France}
\affiliation[c]{Department of Astrophysics, University of Vienna, Türkenschanzstraße 17, 1180 Vienna, Austria}

\emailAdd{clement.stahl@unistra.fr}

\abstract{Collisionless simulations of structure formation with significant local primordial non-Gaussianities at Mpc scales have shown that a non-Gaussian tail favouring underdensities, with a negative $f_{\rm NL}$ parameter, can significantly change the merging history of galaxy-sized dark matter halos, which then typically assemble later than in vanilla $\Lambda$CDM. Moreover, such a small-scale negative $f_{\rm NL}$ could have interesting consequences for the cosmological $S_8$ tension. Here, we complement our previous work on collisionless simulations with new hydrodynamical simulations of galaxy formation in boxes of 30 Mpc/$h$, using the {\sc RAMSES} code. In particular, we show that all feedback prescriptions being otherwise identical, simulations with a negative $f_{\rm NL} \sim -1000$ on small scales, hence forming galaxies a bit later than in vanilla $\Lambda$CDM, allow to form simulated galaxies with more disky kinematics than in the vanilla case. Therefore, such small-scale primordial non-Gaussianities could potentially help alleviate, simultaneously, tensions in cosmology and galaxy formation. These hydrodynamical simulations on small scales will need to be complemented with larger box simulations with scale-dependent non-Gaussianities, to statistically confirm these trends and explore their observational consequences in further detail.}

\begin{document}
\maketitle
\flushbottom

\section{Introduction}\label{sec:Introduction}

The matter content of the $\Lambda$CDM cosmological model is dominated by cold dark matter (CDM) which forms structures out of primordial nearly-scale invariant Gaussian perturbations arising out of the epoch of inflation. Deviations from such Gaussian initial conditions are strongly constrained on large scales \cite{Planck:2019kim, Mueller:2021tqa}, but this is not the case on scales of $\sim \mathcal{O}(10)$ Mpc/$h$.

In view of various tensions and challenges for galaxy formation in the $\Lambda$CDM framework \cite{Abdalla,Bullock}, for instance, the fact that simulated galaxies tend to have too massive bulges and/or too massive stellar halos compared to observed ones \cite{Peebles:2020bph}, the difficulties to form the right bar fraction \cite{Reddish,Roshan} and properties, or the phase-space correlation problem of satellites around galaxies in the local Universe \cite{Pawlowski1, Ibata:2014csa, Muller}, it was recently shown \cite{Stahl:2022did} that local primordial non-Gaussianities (PNG) on scales of $\sim \mathcal{O}(10)$ Mpc/$h$ could potentially alleviate some of these tensions.

Primordial non-Gaussianities \cite{Acquaviva:2002ud} are a smoking gun of non-trivial physics during inflation. To date, significant efforts have been made to assess the impact of PNG on large-scale structures \cite{Achucarro:2022qrl}. Fewer groups focused on the impact of PNG on galactic scales \cite{Avila-Reese:2003cjm,MoradinezhadDizgah:2013rkr,Habouzit:2014hna,Chevallard:2014sxa,Sabti:2020ser,Biagetti:2022ode,CMB-S4:2023zem}. Non-Gaussian initial conditions with skewness are traditionally characterized by a parameter $f_{\rm NL}$ which is the amplitude of the quadratic correction to the Gaussian random field. In Ref.~\cite{Stahl:2022did}, we have performed dark matter-only (DMO) collisionless simulations \cite{Angulo:2021kes} in a box of 30 Mpc/$h$ featuring four different types of primordial non-Gaussianities, among which two models were implemented with non-zero skewness. The model with a positive skewness in terms of the gravitational potential, which was dubbed NG1$+$, translates into a large tail for negative contrast density (hence a negative $f_{\rm NL}$ parameter), whilst the case with negative skewness in terms of the potential, which was dubbed NG1$-$, translates into a large tail towards large positive overdensities (hence a positive $f_{\rm NL}$ parameter).

We found that non-Gaussian models display a distinct and potentially detectable feature in the matter power spectrum around the nonlinear scale $k \sim 3$ $h$/Mpc. The feature is particularly interesting in our NG1$+$ model with negative $f_{\rm NL} \sim -1000$, as the decrease in power compared to the Gaussian case may be very relevant to backup a potential solution to the $S_8$ tension \cite{Amon:2022azi}. We also noted that the UV galaxy luminosity function derived from the Hubble Space Telescope observations had also previously provided a tentative constraint showing that when non-Gaussianities are only present at scales smaller than $\sim 6$~Mpc, the best fit was $f_{\rm NL} \approx -1000$, hence almost exactly like our NG1$+$ model, with a departure from $f_{\rm NL} =0$ significant at $1.7 \sigma$ \citep{Sabti:2020ser}.

On the other hand, we found out that the model NG1$-$ with positive $f_{\rm NL} \sim 1000$, displays possible signs of kinematic coherence of subhalos orbiting around Milky Way-like halos, thereby possibly alleviating the satellite phase-space correlation problem of $\Lambda$CDM \cite{Muller}. We also pointed out that this model displayed a faster mass assembly, with a quieter history and a more empty environment at late times, which we considered to be promising in view of alleviating potential tensions with unexpectedly massive galaxies at high redshift observed with JWST \cite{Labbe, MBK2}, and which we speculated could also help alleviate the tension on the bulge fraction in simulated galaxies, potentially associated with a too violent merging history in $\Lambda$CDM simulations \cite{Peebles:2020bph}. 

In Section~\ref{sec:DMO}, we first complement the 3 simulations featuring $f_{\rm NL} \simeq \{-1000, 0, 1000\}$  performed in our previous work in boxes of (comoving) 30 Mpc/$h$ with $3 \times 20$ collisionless simulations at the same scale. Running these additional collisionless simulations allows us to take into account different initial random seeds, in order to benefit from larger statistics for the shape of the non-linear power spectrum. The next pressing question we want to answer is how all the previously detected effects for dark matter halos translate when one considers hydrodynamical simulations of galaxy formation in this context. In Section~\ref{sec:NGIC}, we describe how we set up 3 hydrodynamical simulations with the {\sc RAMSES} code, based on the 3 simulations of our previous work. Then, we explore in Section~\ref{sec:disk} some baryonic properties of the galaxies that have been formed in this context. Based on four diagnostics (the rotational vs.~dispersion velocities of stars, the in-situ vs.~ex-situ fraction of formed stars, the specific star formation rate sSFR and the $M_{\rm star}$-$M_{\rm halo}$ relation), we show that contrary to our initial guess \cite{Stahl:2022did}, the model with {\it negative} $f_{\rm NL} \sim -1000$ tends to form, all things being otherwise equal, more disky galaxies than in the Gaussian case, while the trend is reversed for the NG1$-$ model with positive $f_{\rm NL} \sim 1000$. This means that the local PNG model which might have interesting consequences for the $S_8$ tension is also potentially promising in view of forming more disky galaxies.

\section{A suite of collisionless simulations}
 \label{sec:DMO}

In this section, we statistically compare the results presented in Ref.~\cite{Stahl:2022did} for the Gaussian and for 2 non-Gaussian simulations with skewness, NG1$+$ and NG1$-$ (hereafter denoted NG$+$ and NG$-$ for simplicity), by simulating $3 \times 20$ times the same $30$ Mpc/$h$ comoving box in DMO, but varying the initial random seed. We apply the exact same procedure to generate the initial conditions as in Ref.~\cite{Stahl:2022did}, to which we refer the reader for the details. We found out that the only quantity computed in Ref.~\cite{Stahl:2022did} that varies significantly is the power spectrum. The other quantities display a qualitative agreement with the original random seed. 

   \begin{figure}
\includegraphics[width=8 cm]{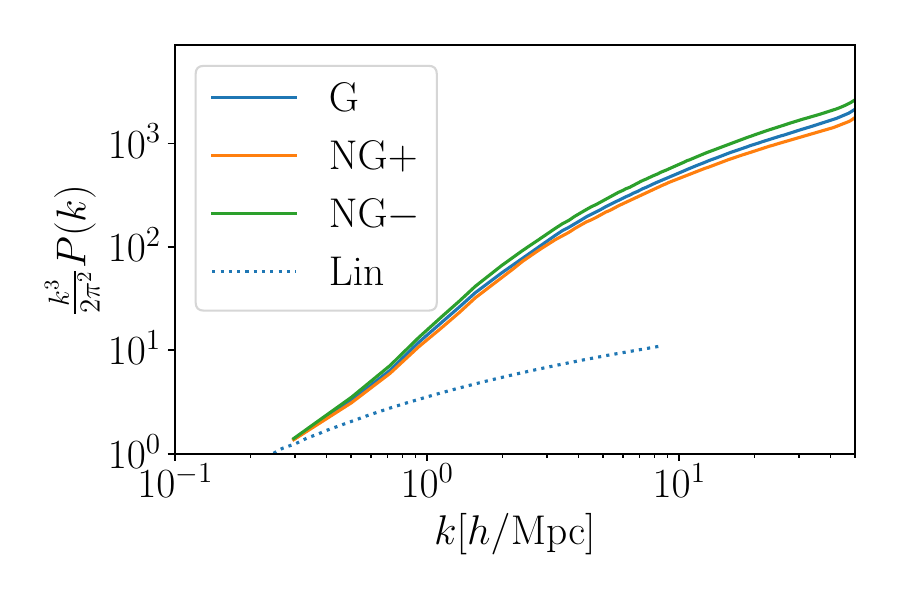} 
\includegraphics[width=8 cm]{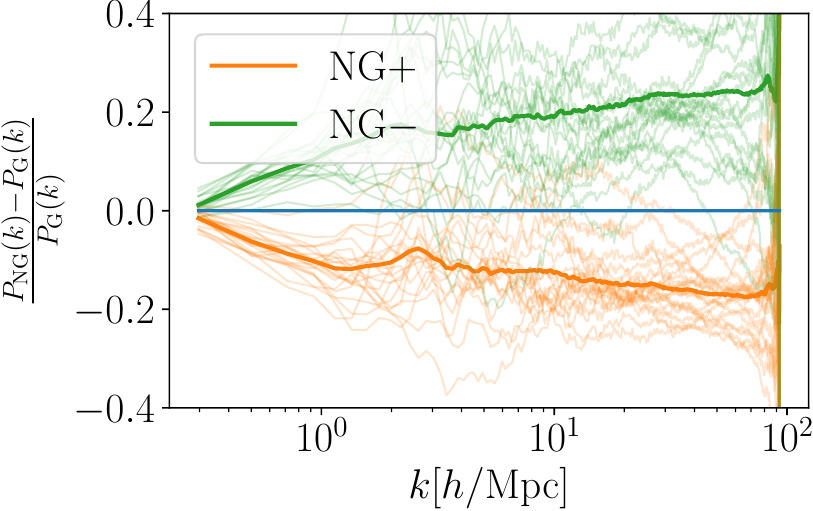}
\caption{\label{fig:PS} Left panel: Average over the 20 realisations of the Gaussian (G) and non-Gaussian (NG+ and NG$-$) dimensionless power spectra at $z=0$, compared to a linear prediction from \texttt{CLASS} \cite{Blas:2011rf} (dotted line). Right panel: Ratio of the DMO power spectra with Gaussian and non-Gaussian initial conditions. The thick line represents the mean over the 20 random seeds. All our results are displayed at $z=0$.}
         \end{figure}

We present in Fig.~\ref{fig:PS} our results for the dimensionless power spectrum at $z=0$ for the NG$+$ and NG$-$ simulations in the nonlinear regime, as well as the relative differences with the Gaussian case. There are variations resulting from the different random seeds\footnote{Erratum: Note that the power-spectrum of figure 4 at $z=0$ in Ref.~\cite{Stahl:2022did} had a bug, which has been corrected here. Moreover, looking at one power-spectrum in only one small box of 30 Mpc/$h$ is not very relevant, due to the variance related to the random seed, shown in Fig.~\ref{fig:PS}.}, but the average trend for NG$+$ and NG$-$ is clear: in particular, the NG+ simulations, starting from primordial non-Gaussianities with a skewness favouring large negative density contrasts (hence with negative $f_{\rm NL} \sim -1000$), generate on average power spectra at $z=0$ that are $10$ to $20\%$ lower than the Gaussian case. This could be of particular interest in view of the so-called $S_8$ tension in cosmology \cite{Abdalla}. The $S_8 = \sigma_8 \sqrt{\Omega_m/0.3}$ parameter, where $\sigma_8$ is  the rms of mass fluctuations within 8 $h^{-1}$~Mpc spheres and $\Omega_m$ the matter density parameter, is indeed measured by weak-lensing surveys to be lower than expected according to $\Lambda$CDM. It has recently been argued in Refs.~\cite{Amon:2022azi,Preston:2023uup} that this value of $S_8$ could however be reconciled with $\Lambda$CDM if the power spectrum is suppressed on nonlinear scales by $10$-$30\%$ compared to the DMO $\Lambda$CDM simulations, hence more suppressed than usually assumed in analyses of weak lensing surveys. Such suppression of the power spectrum could be related to feedback from the baryons, but no current simulation reproduces the necessary suppression between redshift $z=0$ and $z=1$. Hence, the suppression of $10$ to $20\%$ of the nonlinear power spectra from local PNG alone in our DMO simulations with negative $f_{NL} \sim -1000$ on 30 Mpc/$h$ scales might provide an interesting alternative or complement to the feedback mechanisms invoked to suppress the power spectrum. We leave for future work to consider scale-dependent non-Gaussian initial conditions and to gauge its impact on the power spectrum and on the $S_8$ tension. For the rest of this work, we investigate the consequences on galaxy formation with hydrodynamical simulations. Since such simulations are computationally expensive, we consider only our 3 original random seeds for the simulations G, NG+, and NG$-$ in the presence of baryons.

 \section{Hydrodynamical setup} 
 \label{sec:NGIC}

We now resimulate the models G, NG+, and NG$-$ from Ref.~\cite{Stahl:2022did} in the presence of baryons. Using {\sc monofonIC} \cite{Michaux:2020yis,Hahn:2020lvr}, the initial perturbations are drawn from the usual power-law power spectrum at the end of inflation and turned non-Gaussian following Ref.~\cite{Stahl:2022did}. Afterwards they are convolved with the total matter transfer function and the resulting total matter is used as input for third-order Lagrangian perturbation theory (3LPT) to propagate the growing mode, while the baryon-CDM isocurvature perturbation and decaying mode (at first order) is imposed as mass variations of the particles/gas cells. Our hydrodynamical simulations are performed with the adaptive mesh refinement code {\sc ramses}~\citep{Teyssier:2001cp}. 
The box size is $30 \, {\rm Mpc}/h$, and a $\Lambda$CDM cosmology compatible with the Planck data is assumed, with total matter density parameter $\Omega_{\rm m}=0.31$, dark energy density parameter $\Omega_\Lambda=0.69$, baryon matter density $\Omega_{\rm b}=0.0455$,  Hubble constant $H_0=67.7\, \rm km\,s^{-1}\, Mpc^{-1}$,  primordial amplitude $A_s=2.1 \times 10^{-9}$, and spectral index of $n_{\rm s}=0.968$.
The dark matter mass resolution of the simulations is $M_{\rm DM,res}=2.2 \times 10^{7}\, \rm M_\odot$.
We allow for several levels of refinement up to level 15, where final levels are activated at expansion factors of $0.2,0.4,0.8$ so as to keep an approximately constant minimum cell size resolution in proper units of $\Delta x \simeq 1\, h^{-1}\,\rm kpc$.
The different levels are refined, respectively derefined, everywhere the total mass within a cell is larger, respectively lower, than 8 times the dark matter mass resolution.
Particle dynamics is followed using a standard leap-frog algorithm using the gravitational acceleration on the grid obtained with a particle-mesh method.
Hydrodynamics is evolved on the mesh with a second-order MUSCL-Hancock method using a total variation diminishing scheme with minmod slope limiter and the HLLC Riemann solver.
Gas is monoatomic with an initial primordial composition of 76 \% of hydrogen and 24 \% of helium, and follows an equation of state with an adiabatic index of $5/3$.

 The main simulations closely follow the baryonic physics of Horizon-AGN \cite{Dubois:2014lxa,Dubois_2016}. We summarize here their main properties.
 Gas can cool down to a minimum temperature of $T=10^4\,\rm K$ due to H, He, and metals produced by SNe, following~Ref.~\cite{Sutherland:1993ec} within a uniform \cite{Haardt:1995bw} UV background heating after reionisation at redshift $z=10$.
The gas pressure at densities larger than $n_0=0.1\,\rm H\, cm^{-3}$ follows a polytropic equation of state with polytropic index $4/3$.
Star formation follows a Schmidt relation with a low star formation efficiency of $\varepsilon_{\star}=0.02$ for gas with density larger than $n_0$.
New star particles of mass resolution $M_{\star,\rm res}=1.3\times 10^7\,\rm M_\odot$ are formed following a Poisson sampling process \cite{Rasera:2005gq}.
After 10~Myr of evolution, a  star particle releases all the energy, gas (and metals) of its core-collapse (type II) supernovae (SNII) back to the gas component. 
For the fiducial model, we assume that the specific energy released by SNII is $e_{\rm II,fid}=3\times 10^{49}\,\rm erg\, s^{-1}$, that the fraction of mass returned is $f_{\rm mII,fid }=30 \%$ and that the metal yield is $f_{\rm ZII,fid}=5\%$.
Black holes (BH) with seed mass of $10^5\,\rm M_\odot$ form in gas densities above $n_0$ if there is no other BH closer than 50 kpc (in order to avoid forming multiple BHs in the same galaxy).
Two BHs can merge if they are closer than $4\Delta x$ from each other.
BHs are modeled as sink particles and can grow by gas accretion following of Bondi-Hoyle-Littleton accretion rate limited at Eddington, assuming a standard radiative efficiency of $\varepsilon_{\rm r}=0.1$.
This gas accretion onto BH leads to active galactic nuclei feedback with a dual mode depending on the Eddington ratio $\chi_{\rm E}$ (accretion rate over Eddington rate): a jet/kinetic mode if $\chi_{\rm E}<0.01$ and quasar/heating mode otherwise with feedback efficiencies calibrated to reproduce the low redshift BH-to-galaxy mass scaling relations. 

It is useful to keep in mind that our goal in the following is to isolate the effects of PNG on galactic scales by comparing simulations with the exact same subgrid recipes. However the fiducial feedback is actually calibrated\footnote{More precisely, the subgrid recipes encompass a complete galactic physics model that allows the simulations to be within the bounds on a set of observed quantities \cite{Dubois_2016}.} on vanilla $\Lambda$CDM. Hence, there could be a potential need for a recalibration to the extreme small-scale PNG considered here. In order to have information on the possible degeneracies between baryonic physics and PNG, we therefore also explore two other sets of simulations for the three cases G, NG+, and NG$-$: one with the AGN feedback turned off (dubbed ``NoAGN''), and one with the AGN feedback turned off and a very low input from SNII using $e_{\rm II,fid}=5\times 10^{48}\,\rm erg\, s^{-1}$, $f_{\rm mII,fid }=5 \%$ (dubbed ``NoAGNNoSn''). We thus run a total of 9 hydrodynamical simulations, all starting at $z=50$. These experiments will not fully address whether a recalibration of the fiducial feedback would be needed in the non-Gaussian cases, but they will shed light on the possible degeneracies.

\section{Results}
\label{sec:disk}

Simulations were run down to redshift $z=0$, and the analysis is performed with 56 snapshots between $z=15$ and $z=0$. All error bars in this paper are computed using the standard error on the mean. Halos and galaxies were identified using the finder {\sc AdaptaHOP} \cite{Aubert:2004mu}. We consider in the following only galaxies with a stellar mass of at least $4 \times 10^{8}\, \text{M}_{\odot}$ corresponding to $\sim$ 40 star particles, meaning that the choice of halo finder does not matter for the results reported. 

\begin{figure}[ht]
    \includegraphics[width=\textwidth]{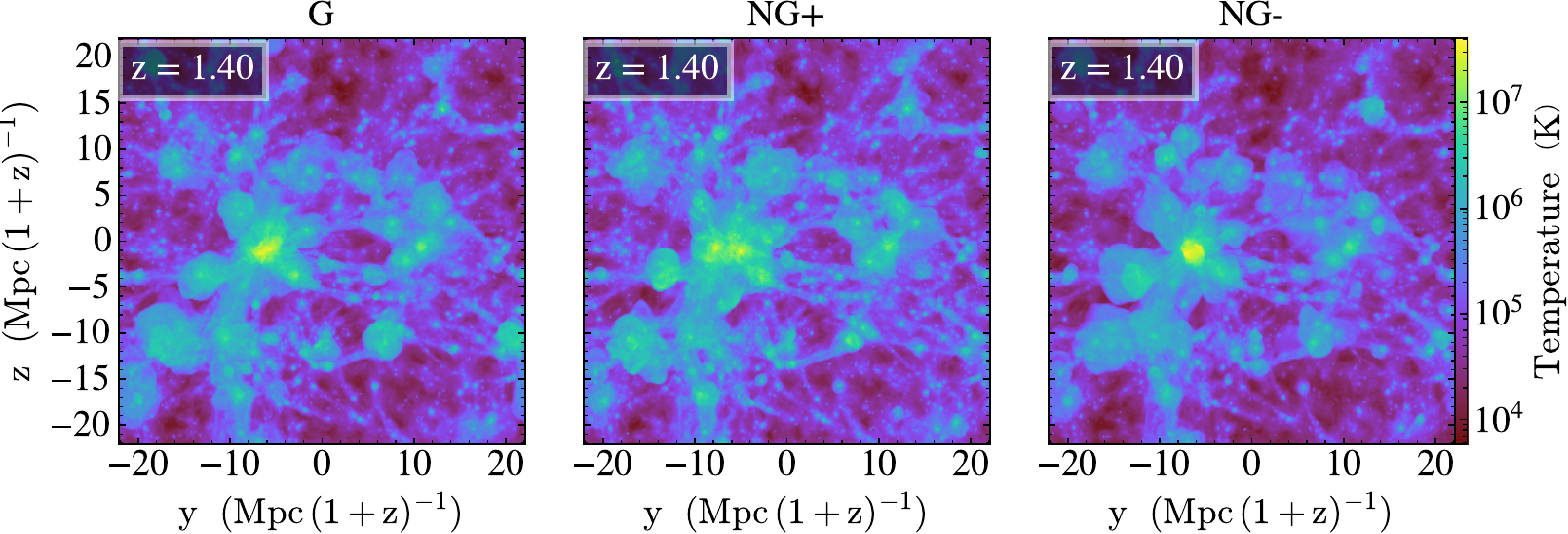}        \caption{\label{fig:visu} Visualization of the temperature of the gas at redshift $z=1.4$.}
\end{figure}

Fig.~\ref{fig:visu} displays a visualization of the gas temperature in the simulation at redshift $z=1.4$ for the Gaussian and NG$\pm$ simulations in the fiducial feedback case. Having larger overdensities initially (NG-) leads to a more pronounced projection map than the Gaussian case, with more clumped structures and more empty regions in between. Having more underdense regions (NG+), the structures and voids form later than in the Gaussian case, resulting in a smoother temperature projection map.

\subsection{Disk kinematics}
\label{sec:vsigma}

\begin{figure}[ht]
    \includegraphics[width=\textwidth]{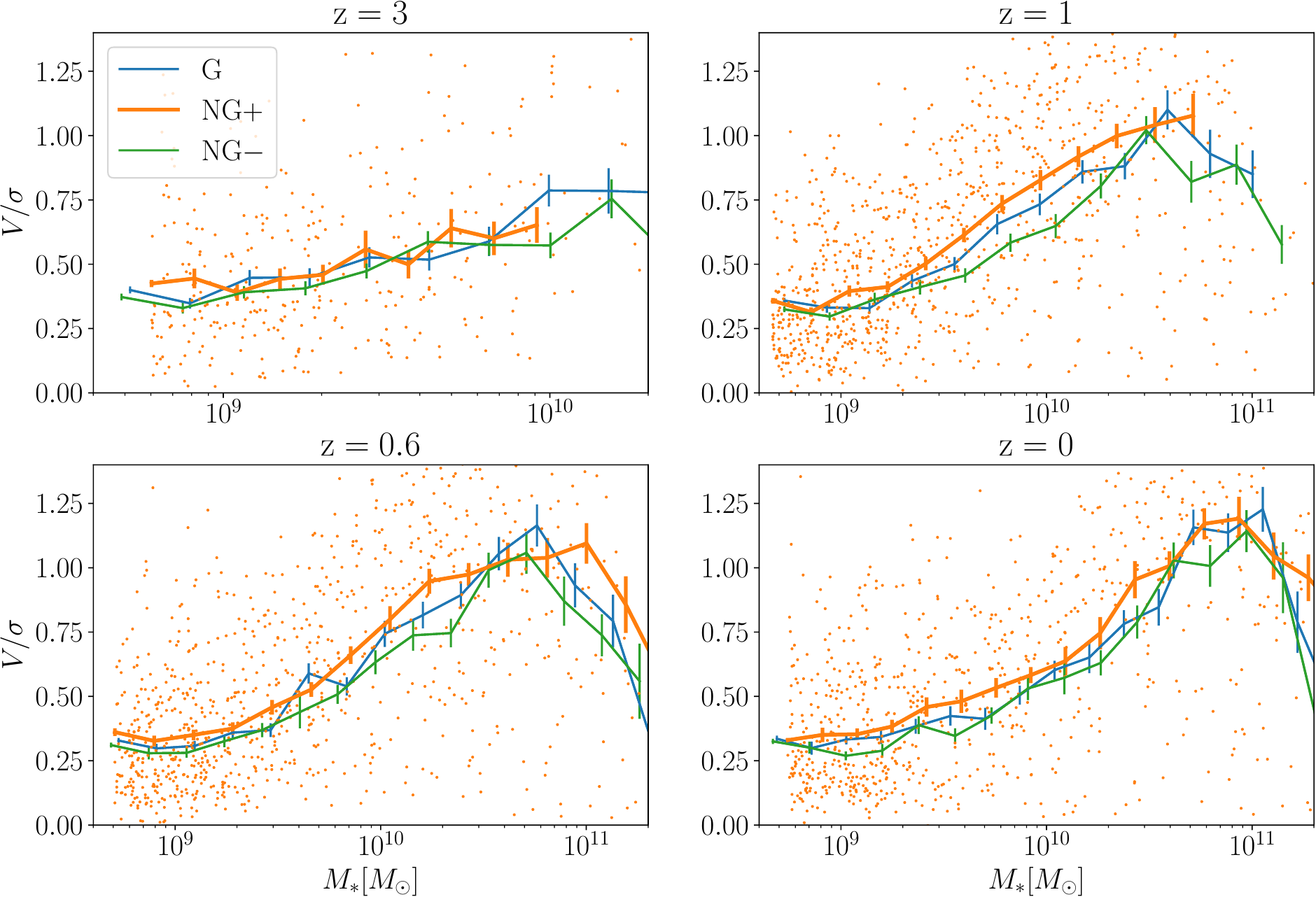}        \caption{\label{fig:quadri} For different redshift, $V/\sigma$ as a function of the galaxy mass. The orange dots represent the NG+ measure for each galaxy, we only displayed it to avoid excessive clutter.}
\end{figure}

As a first indicator of the morphology of galaxies, we use the kinematic ratio of their rotation- to dispersion-dominated velocity $V/\sigma$ (Figure \ref{fig:quadri}). This quantity is computed from the velocity distribution of stellar particles of each galaxy: the tangential velocity is computed in the plane perpendicular to the angular momentum of the stellar component of each galaxy. We selected only stars encompassed in a sphere of radius $2 \times R_{\rm eff}$, where the effective radius $R_{\rm eff}$ of each galaxy is defined as in Ref.~\cite{Dubois_2016}.
Note that $V/\sigma$ is not directly comparable to observational measures \cite{Jang} but is an indicator of the diskiness of the galaxies when comparing simulations with respect to each other.

\begin{figure}[ht]
    \includegraphics[width=\textwidth]{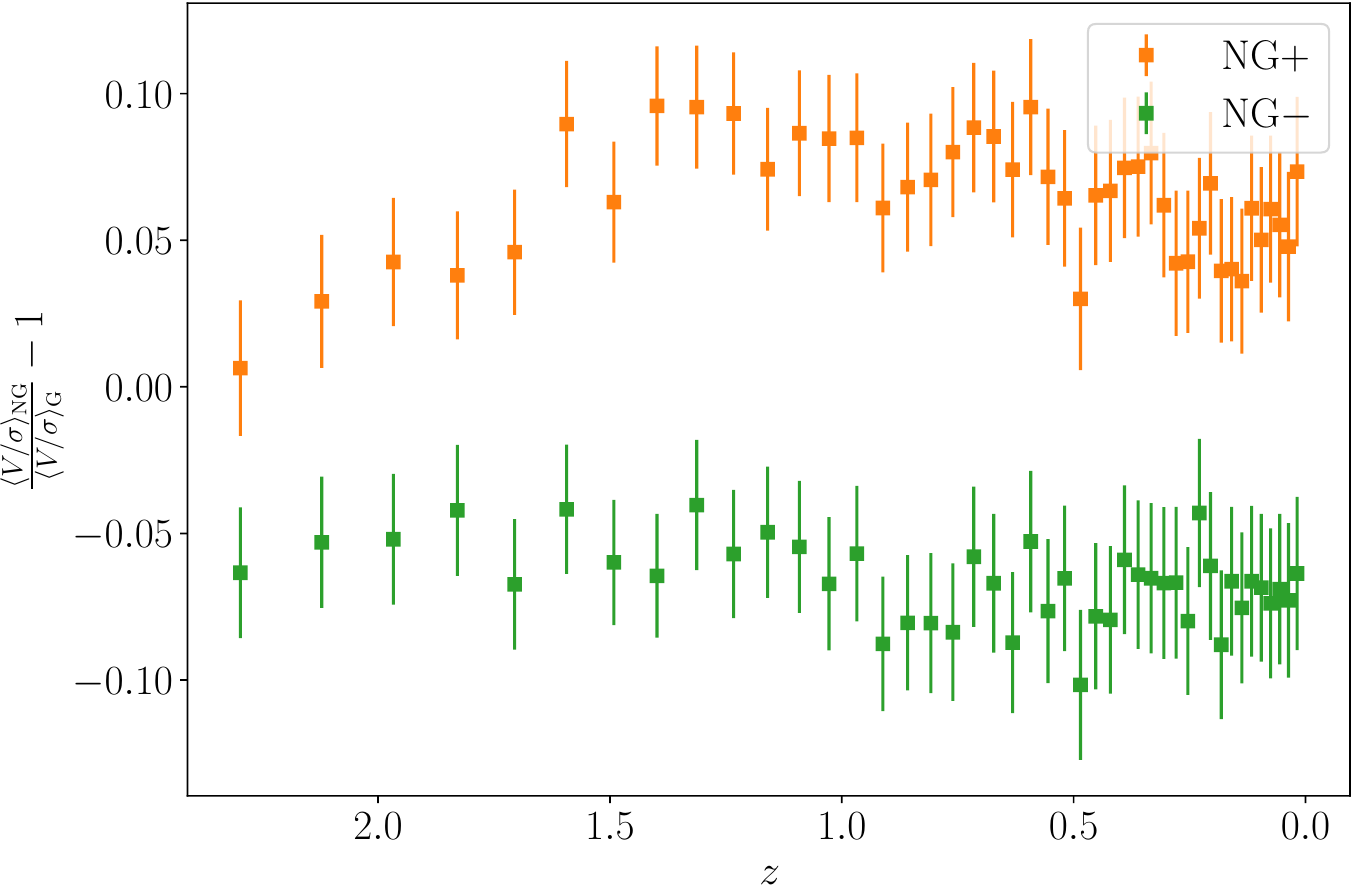}        \caption{\label{fig:quadri2} Mean $V/\sigma$ of the whole samples of galaxies for the non-Gaussian models divided by its Gaussian counterpart as a function of redshift. There is a sizable difference of 5-10 \% between the models.}
\end{figure}

The evolution of $V/\sigma$ as a function of redshift is shown in Fig.~\ref{fig:quadri2}.
 As can clearly be seen, the effect of PNG is mild but significant :
$
\langle (V/\sigma)_{NG-} \rangle < \langle (V/\sigma)_G \rangle <\langle (V/\sigma)_{NG+} \rangle ,
$
at all redshifts, and at essentially all stellar masses. The differences between these average ratios of rotation- to dispersion-dominated velocity in the Gaussian and non-Gaussian cases are of the order of $5-10$\% and much smaller than the intrinsic dispersions of the individual values of the $V/\sigma$ values. For the model initially favouring a tail of large underdensities, NG+, galaxies are typically {\it more disky} than in the vanilla Gaussian case. At $z=0$, the average $\langle V/\sigma \rangle$ of all NG+ simulated galaxies is $7.4\%$ higher than in the Gaussian case.

\subsection{In-situ vs ex-situ stars}
\label{sec:insitu}

\begin{figure}[ht]
 \includegraphics[width=\textwidth]{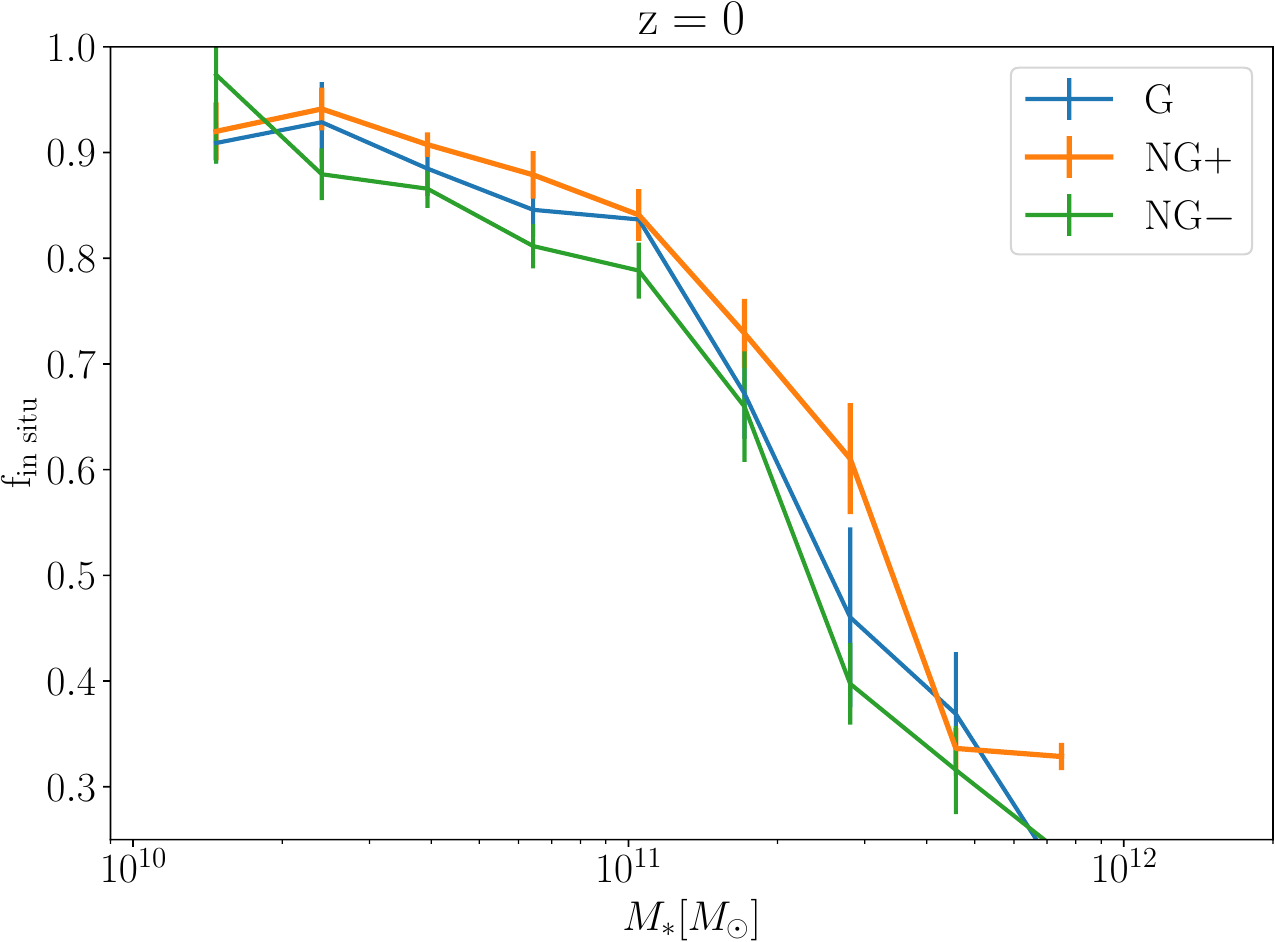} \caption{\label{fig:IS} Mean fraction of in situ formed stars as a function of stellar mass at $z = 0$. }
\end{figure}

An interesting quantity to compute to understand this slight preference for diskiness of the NG+ simulation compared to the Gaussian case (and even more compared to the NG$-$ case) is the fraction of stars formed in-situ and ex-situ at redshift $z=0$. As a reminder, the NG$-$ simulation favours overdensities which grow fast at high redshift before having a calmer environment at later times. Intuitively, one might have expected that this calmer environment at later times would have favoured a larger fraction of stars formed in situ and more disky geometries, but the contrary happens, as shown in Fig.~\ref{fig:IS}.

The active early phases of star formation and violent early merger history in NG$-$ brings many ex-situ stars in the galaxies formed in this scenario, and the specific star formation rate later decreases compared to the Gaussian case, meaning that the fraction of ex-situ formed stars is larger than in the Gaussian case and that the galaxies end up being slightly less disky on average. The contrary happens for the NG+ case initially favouring a tail of large underdensities, in which galaxies end up slightly more disky and with more stars formed in situ than in the Gaussian case.

\subsection{Specific star formation rate and $M_{*}$-$M_{h}$ relation}
\label{sec:ssfr}

\begin{figure}[ht]
 \includegraphics[width=\textwidth]{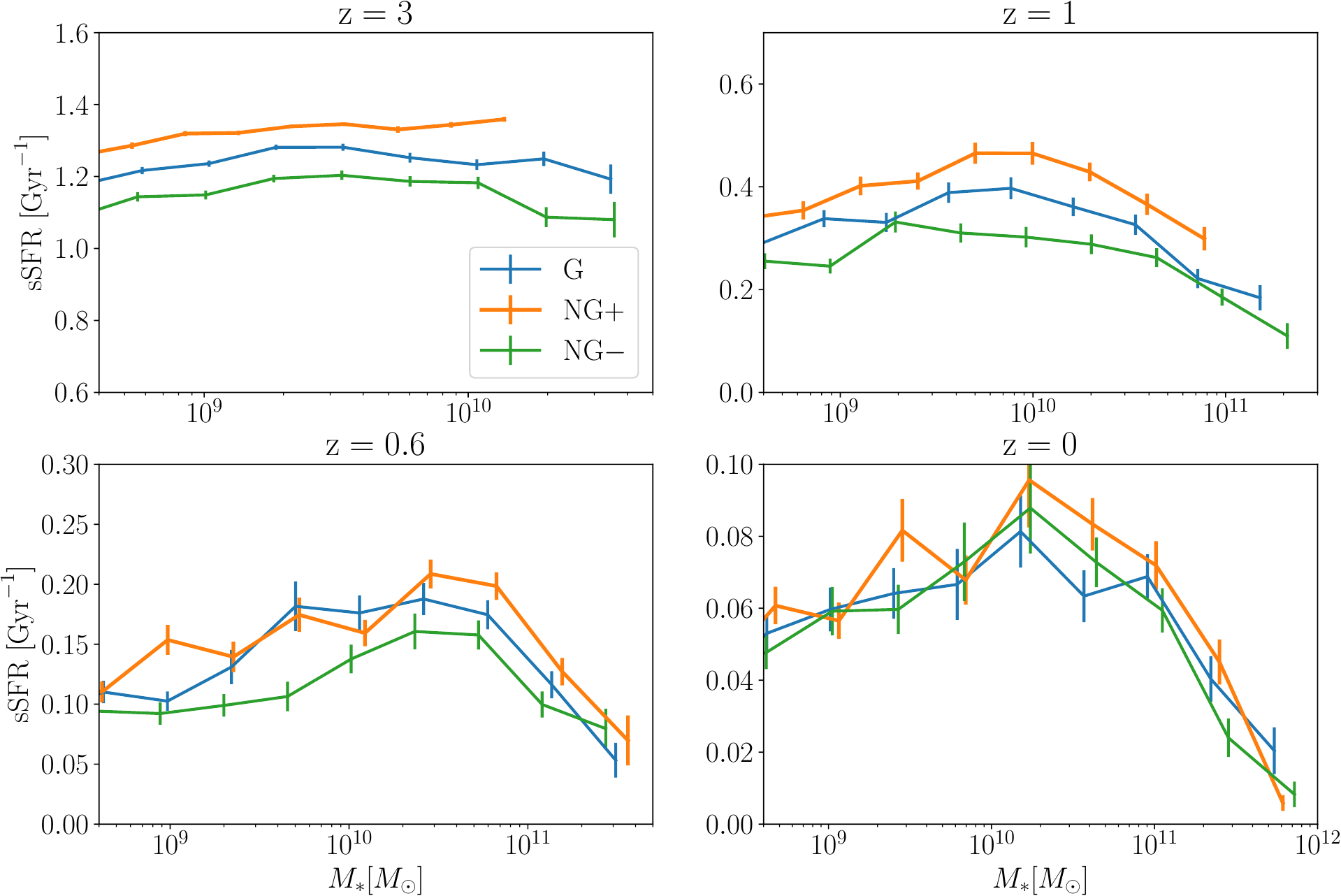} \caption{\label{fig:sSFR} Specific star formation rate as a function of stellar mass at different redshifts. Note the different scales for the y-axis. At high redshift, there is a clear trend that the model NG+ forms more stars. It is consistent with the fact that structure formation starts later in that model.}
\end{figure}

To confirm the above scenario, we display in Figure \ref{fig:sSFR} the specific star formation rate ${\rm sSFR} = {\rm SFR}/M_*$ which is calculated for each snapshot over its past Gyr. As expected from the discussion in the previous subsection, NG$-$ started its structure formation earlier and therefore, all other parameters being identical, has a lower sSFR than the Gaussian case at all redshifts smaller than 3. The Gaussian case itself has a lower sSFR than NG+. This higher star formation rate on average in the NG+ case is the reason for the slightly more disky geometry on average, as quantified by the $V/\sigma$ parameter.

\begin{figure}[ht]
 \includegraphics[width=\textwidth]{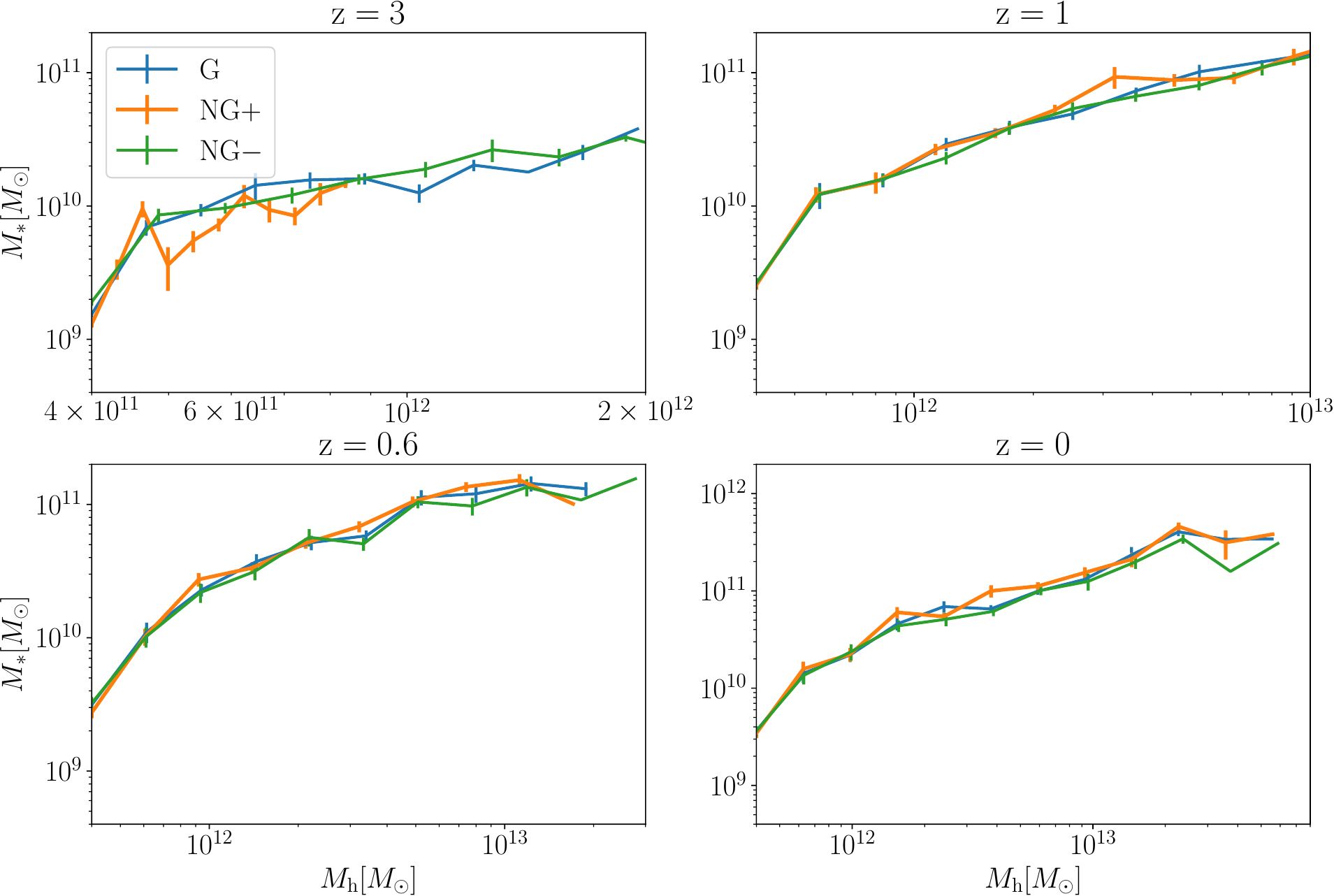} \caption{\label{fig:MhMs} The stellar-to-halo mass relation for different redshifts. As the model NG+ builds its galaxies later, at redshift 3, it has fewer stars at fixed halo mass but it catches up at later times to slightly dominate at redshift 0.}
\end{figure}

In Figure \ref{fig:MhMs}, we display the stellar-to-halo mass relation of the different simulations. This relation was also displayed in Figure 10 of Ref.~\cite{Dubois_2016} in a larger box for the Gaussian case, and was shown to agree with observational constraints on the luminosity function. The differences are minor, but one can notice that, since the NG+ model forms its galaxies later, it has typically fewer stars at a given halo mass at redshift 3. However, over time, it gradually catches up with efficient star formation, and eventually ends up with slightly more stellar mass in galaxies at fixed halo mass.

\subsection{Degeneracies between baryonic physics and non-Gaussian initial conditions}

\begin{figure}[ht]
    \includegraphics[width=\textwidth]{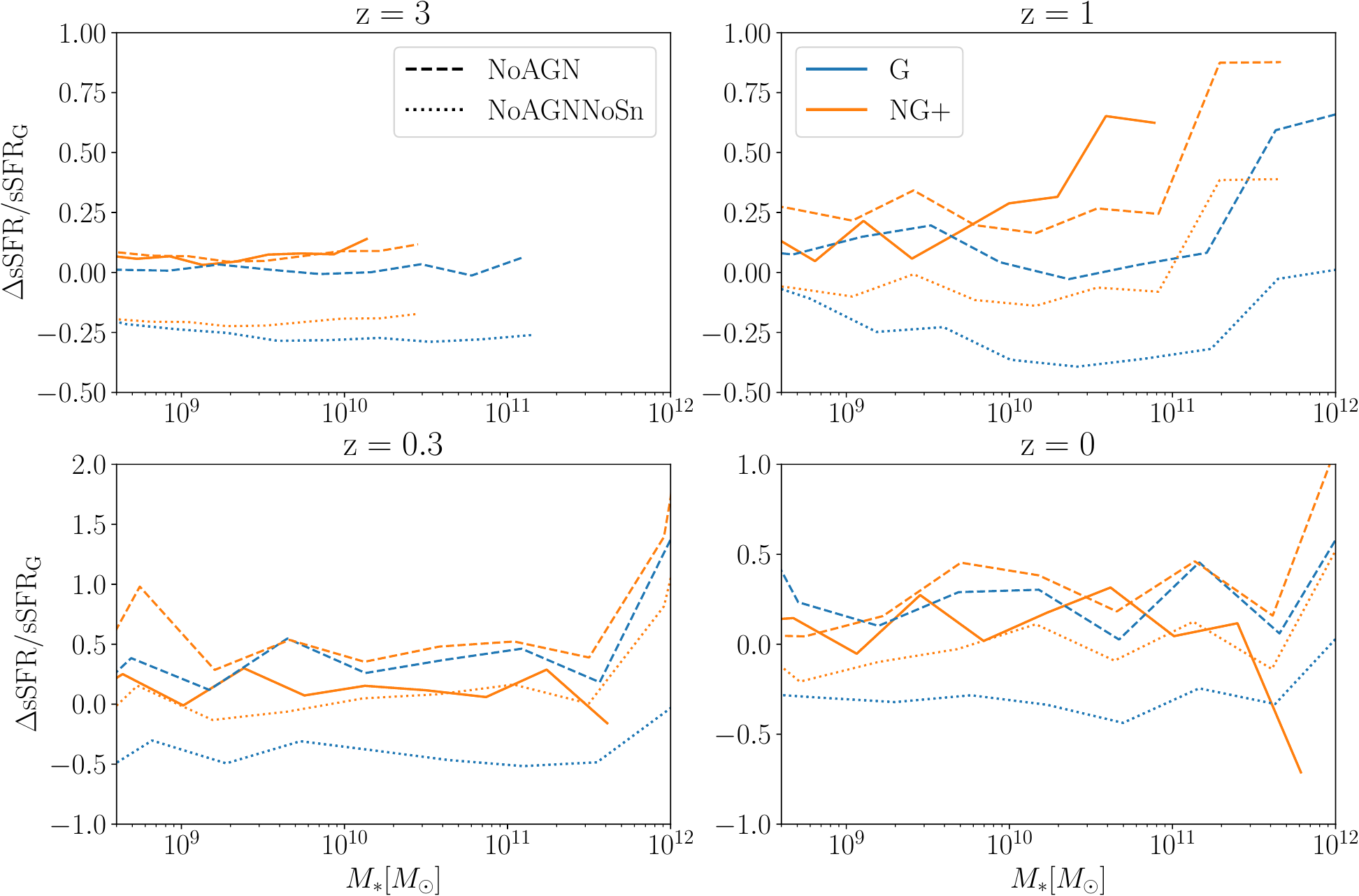}        \caption{\label{fig:sSFR_NoFEED} The relative difference in the specific star formation rate with the Gaussian model G with the feedback prescription of Horizon-AGN as a benchmark. }
\end{figure}

The results obtained in this paper indicate that, {\it all being otherwise equal}, the non-Gaussian model with negative $f_{\rm NL}$, NG+, yields, on average, more disky geometries than the vanilla Gaussian case without changing any feedback recipe. The contrary is true for the NG$-$ simulation. In order to get a flavour of the possible degeneracies between baryonic feedback and this effect of local PNG, we analyze two additional sets of simulations without AGN feedback (dubbed ``NoAGN'') and without both AGN feedback and supernovae feedback (dubbed ``NoAGNNoSn''). As an illustrative example, we calculated the specific star formation rate calculated over the last Gyr for each scenario. The results are plotted for G and NG+ on Fig.~\ref{fig:sSFR_NoFEED}.

We note that the presence of AGN feedback in the simulation results in a slight decrease of the sSFR at all masses and a strong decrease at high mass ($M_{\rm stars} > 10^{11} M_{\odot}$). Therefore, turning off the AGN feedback slightly increases the sSFR at all masses compared to the fiducial Gaussian case. This is in line with what happens in NG+, but the two effects are very different at high masses, where the sSFR increases much more strongly than in NG+ when turning off the AGN feedback. On the other hand, the supernova feedback tends to boost the formation of stars at low mass ($M_{\rm stars} < 10^{11} M_{\odot}$), hence turning off SNs tends to strongly decrease the sSFR. In the NG+ model, this results in an sSFR comparable to the fiducial case at $z<1$, but a much lower one at $z=3$. Shutting off both the input from SN II and AGN feedback then also boosts the sSFR at high masses, where the shutting off of AGN dominates. It is thus clear that some of the effects detected in this work are at least partially degenerate with the implementation of baryonic physics, but it is not clear that they are fully degenerate when considering the effect at all redshifts. Local PNGs can in any case mimic or increase some trends caused by changing the feedback recipes without making those recipes less realistic.

\section{Conclusions}
\label{sec:concl}

In Ref.~\cite{Stahl:2022did}, we found some interesting potential signatures of small-scale local PNGs on the formation of galaxy-sized dark matter halos in collisionless simulations. Here, we further investigated the impact of primordial non-Gaussianities on galaxy formation through hydrodynamical simulations including baryonic feedback. We closely follow the prescription for baryonic feedback used in the Horizon-AGN simulation \cite{Dubois_2016}.

We focused on the simulations including skewness in the primordial distribution of contrast densities, NG+ ($f_{\rm NL} \simeq -1000$) and NG- ($f_{\rm NL} \simeq 1000$). We first complemented these collisionless simulations on 30 Mpc/$h$ scales by other ones with different initial random seeds, in order to benefit from larger statistics for the shape of the nonlinear power spectrum. The NG+ simulations generate on average power spectra at $z=0$ that are 10 to 20\% lower than the Gaussian case, which could be of particular interest in view of the $S_8$ tension. Indeed, it has been argued to be fully reconcilable with $\Lambda$CDM if the power spectrum is suppressed on nonlinear scales by 10-30\% compared to collisionless $\Lambda$CDM simulations \cite{Preston:2023uup}. The effect of small-scale local PNG in combination with baryonic feedback could therefore help reach this goal without making the feedback unrealistic.

We then set up 3 hydrodynamical simulations with the {\sc RAMSES} code, based on the 3 simulations of our previous work, and show that all things being otherwise equal, the galaxies formed in NG+ are more disky (rotational vs. dispersion velocities of stars), with more stars formed in-situ and a larger specific star formation rate. This means that the local PNG model, which might have interesting consequences for the $S_8$ tension, is also potentially promising in view of forming more disky galaxies in simulations, which has long been known to be a potential problem of current simulations, also in view of the formation of bars \cite{Reddish}.

Our results therefore show that small-scale local PNG could have a profound impact on galaxy formation: it is a largely unexplored avenue that is motivated by fundamental physics that could shed new light on the field. It will now be mandatory to devise large-box simulations with scale-dependent local primordial non-Gaussianities passing the large-scale constraints: these could then serve as a basis for zoom-in simulations using the same baryonic physics as the one used here, to check whether a deep connection between the physics of inflation and galaxy formation might
exist, and alleviate the $S_8$ tension at the same time.

\acknowledgments
CS, BF, and RI acknowledge funding from the European Research Council (ERC) under the European Union's Horizon 2020 research and innovation program (grant agreement No.\ 834148). TM and OH acknowledge funding from the ERC under the European Union’s Horizon 2020 research and innovation program, Grant Agreement No. 679145 (COSMO-SIMS).
This work has made use of the Infinity Cluster hosted by the Institut d'Astrophysique de Paris.

\section*{Tools}
The analysis was partially made using  \href{https://github.com/franciscovillaescusa/Pylians}{{\sc Pylians}}, \href{https://yt-project.org/}{YT} \cite{Turk:2010ah}, as well as IPython \cite{Perez:2007emg}, Matplotlib \cite{Hunter:2007ouj} and NumPy \cite{vanderWalt:2011bqk}.

\section*{Authors' Contribution}
The simulations presented in this work are extensions of Horizon-AGN (PI: YD). They
were performed and analysed by CS and YD, in consultation with BF. CS and BF drafted the manuscript. All the authors improved it by their
comments.

\section*{Carbon Footprint}
Following Ref.~\cite{berthoud} to convert\footnote{Including the global utilisation of the cluster and the pollution due to the electrical source, the conversion factor is 4.7 gCO2e/h core} the 3.5 M CPU h used to run the simulations presented in this work gives an impact of 16 tCO2eq.

\bibliographystyle{JHEP.bst}
\bibliography{ref.bib}

\end{document}